\documentclass[11pt]{article} 
\pdfoutput=1

\usepackage{times}
\usepackage[pdftex]{graphicx}
\usepackage{fullpage}

\begin{document}
\newcommand{\ignore}[1]{}

\title{Our Brothers' Keepers: Secure Routing with High Performance
       \thanks{This research was supported by an NSERC Discovery grant.}}

\author{
    Alex Brodsky\\
    University of Winnipeg \\
    Winnipeg, MB, Canada, R3B 2E9\\
    {\tt abrodsky@acs.uwinnipeg.ca}
\and
    Scott Lindenberg\\
    University of Winnipeg \\
    Winnipeg, MB, Canada, R3B 2E9\\
    {\tt slindenb@acs.uwinnipeg.ca}
}

\maketitle
\begin{abstract}
  The Trinity~\cite{BrBr07} spam classification system is based on a
  distributed hash table that is implemented using a structured
  peer-to-peer overlay.  Such an overlay must be capable of processing
  hundreds of messages per second, and must be able to route messages
  to their destination even in the presence of failures and malicious
  peers that misroute packets or inject fraudulent routing information
  into the system.  Typically there is tension between the requirements
  to route messages securely and efficiently in the overlay.

  We describe a secure and efficient routing extension that we
  developed within the I3~\cite{StAdZhShSu04} implementation of the
  Chord~\cite{StMoKaKaBa01} overlay.  Secure routing is accomplished
  through several complementary approaches: First, peers in close
  proximity form overlapping groups that police themselves to identify
  and mitigate fraudulent routing information.  Second, a form of
  random routing solves the problem of entire packet flows passing
  through a malicious peer.  Third, a message authentication mechanism
  links each message to it sender, preventing spoofing.  Fourth, each
  peer's identifier links the peer to its network address, and at the
  same time uniformly distributes the peers in the key-space.

  Lastly, we present our initial evaluation of the system, comprising
  a 255 peer overlay running on a local cluster.  We describe our
  methodology and show that the overhead of our secure implementation
  is quite reasonable.

\vspace*{2ex}
\noindent {\bf keywords:} secure routing, peer authentication, 
                          distributed hash tables
\end{abstract}

\thispagestyle{empty}

\newpage
\setcounter{page}{1}

\section{Introduction}
Systems such as Trinity~\cite{BrBr07}, LOCKSS~\cite{MaRoRoBaGiMu03},
and others are based on distributed hash tables that are implemented
on top of peer-to-peer structured overlays.  These overlays differ
from better known peer-to-peer systems such as BitTorrent in three
fundamental ways.   First, these overlays are closed, meaning that
only authorized hosts may join the overlay.  Second, these overlays
must be secure and function even in the presence of failures, denial
of service attacks, and malicious peers.  Third, performance is
paramount, meaning that each peer in the these overlays must be able
to forward hundreds of messages per second.

Although securing closed overlays seems more manageable than the
task of securing open overlays, the task presents several challenges.
First, identifying, authenticating and authorizing peers and
authenticating the messages that they send is not easy because the
mechanisms must be fault tolerant, allow revocation, and must not
significantly impact performance.  Second, securely routing messages, 
dealing with host and network failures, and most importantly, dealing 
with malicious peers and the fraudulent routing information that they 
inject into the overlay is challenging in itself, let alone without
significantly impacting performance.

As part of the Trinity project~\cite{BrBr07}, we have designed,
implemented, and tested a secure closed overlay based on the
I3~\cite{StAdZhShSu04} Chord~\cite{StMoKaKaBa01} implementation.
Our design comprises a distributed and fault tolerant identification, 
authentication, and authorization mechanism; a key assignment scheme 
that encodes a peer's network location yet ensures that the keys are
uniformly distributed in the key space; a self-policing scheme based 
on groups of local peers; and a form of random routing that ensures 
that no (malicious) peer is a choke-point between any two other peers.

In addition to describing our approaches, we present a performance
evaluation, which was performed on a local cluster that hosted
overlays consisting of 255 peers.  We compare the performance of
our system in ``secure'' and ``insecure'' modes, and show that the
performance penalty for secure operation is acceptable.  

The rest of the paper is organized as follows: Section~\ref{sec:bg}
describes our assumptions and the Chord protocol.  Section~\ref{sec:imp}
describes the three parts of our approach and Section~\ref{sec:eval}
describes our evaluation of the system.  Lastly, Section~\ref{sec:related}
and~\ref{sec:conc} describe related work, and discuss future work.

\section{Preliminaries}\label{sec:bg}
We selected the Chord~\cite{StMoKaKaBa01} structured overlay to
provide lookup services for the Trinity~\cite{BrBr07} system because
Chord has good performance characteristics and provides control over 
the location of peers within the overlay, which makes securing the 
overlay easier~\cite{SiMo02,CaDrGaRoWa02}.  

The Chord~\cite{StMoKaKaBa01} overlay structure assigns each peer
a unique key, $k$, from a $160$-bit key-space and organizes the
peers into a single ring in order of their keys.  The predecessor
and successor of key~$k$ are the keys~$k_p$ and~$k_s$, respectively,
belonging to peers in the ring, such that $k - k_p \bmod 2^{160}$
and $k_s - k \bmod 2^{160}$, respectively, are minimal.  Intuitively,
the peer to whom key~$k$ is assigned is located between its predecessor
and successor, the peers to whom the keys~$k_p$ and~$k_s$ are
assigned.  If a key $k$ is not assigned to a peer in the ring, then
the peer whose key is the successor to $k$ is responsible for the
key.  Consequently, each peer is responsible for all the possible
key values between it and its predecessor.

When a peer joins the ring, it locates its position within the ring
by sending a ``find successor'' request with its own key, $k$, to
a ``well known'' peer that is already in the ring.  The request is
routed to the current predecessor of $k$, whose successor is therefore
also the successor of $k$.  The predecessor replies to the new
peer, informing it of both the successor and itself.  The new peer
then informs the successor and predecessor of its existence and assumes
its location in ring.  Lastly, the peer builds its routing table,
called a finger table.

\begin{figure*}[htb]
  \begin{center}
  \ \includegraphics[scale=0.55]{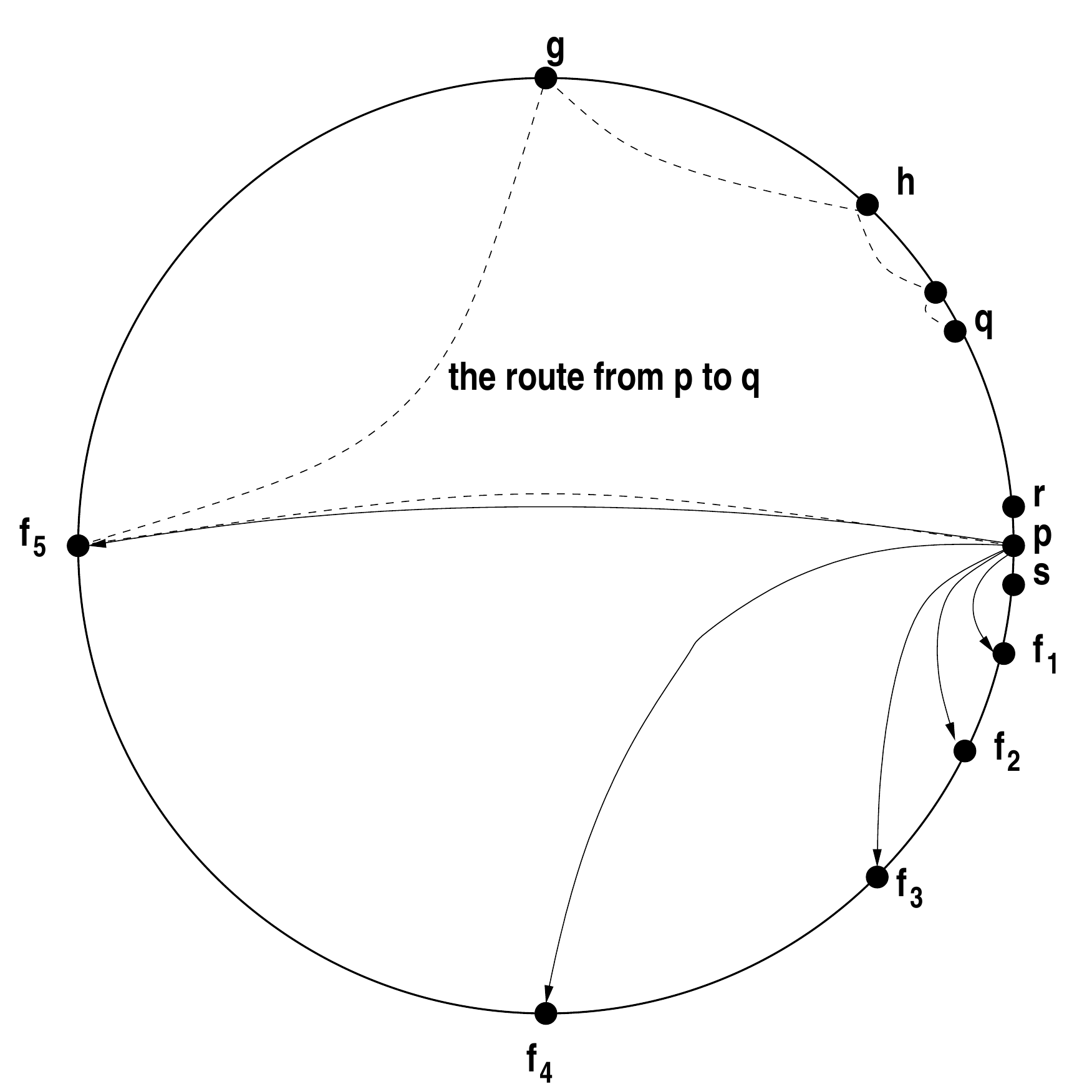}\
  \caption{The peers labeled $f_i$ are in $p$'s finger table, peer~$g$ is 
           in peer~$f_5$'s finger table, and peer~$h$ is in peer~$g$'s 
           finger table.  Peers~$r$ and~$s$ are the predecessor and 
           successor of peer~$p$.  \label{fig:chord}}
  \end{center}
\end{figure*}

The finger table is used by the peer to forward a message toward
its eventual destination.  The finger table comprises keys of select
peers in the ring.  Typically, the table contains $O(\log N)$ keys
of peers that are $\frac{1}{2^i}$ of a ring away, $i = 1 \ldots
\log(N)$, where $N$ is the number of peers in the ring (see Figure
\ref{fig:chord}).   To forward a message to the peer responsible
for key $k$, the peer with the closest preceding key to $k$ is
selected from the finger table, and the message is forwarded directly
to that peer.  Thus, the distance to the destination peer is decreased
by at least half, and after at most $O(\log N)$ such hops, the message
arrives at the destination.  If the closest preceding peer is the
current peer, then the message is forwarded directly to the peer's
successor, its destination.

The finger table is populated by performing additional ``find
successor'' queries with key values of the form $k + 2^i \bmod
2^{160}$, $0 < i < 160$.  Additional ongoing ``find successor''
queries, at regular intervals, are used to update the finger table
as well as the peer's successor and predecessor.  Also, a simple
heart-beat mechanism tracks when peers leave the ring.

Unfortunately, the system as described, is susceptible to many
attacks.  First, the overlay uses an unreliable message-based
transport protocol, User Datagram Protocol (UDP), that is susceptible
to spoofing because the source address of a message can easily be
forged.  Thus, the source of the message can not be (reliably)
determined.  Second, the system, as described, allows any host to
become a peer, which is problematic for a closed overlay and can
lead to the admittance of malicious peers.  Third, as a result of
the first two weaknesses, the overlay is susceptible to denial of
service attacks because large numbers of messages and requests can
be injected into the overlay by external hosts.

Fourth, the overlay relies on the correct behaviour of all of its
constituents.  For example, all peers must correctly forward and
reply to ``find successor'' requests.  Malicious peers can inject
fraudulent routing information into the overlay by replying with
incorrect ``find successor'' replies, dropping requests, or
misdirecting the requests.  Consequently, a few collaborating
malicious peers could cause segments of the ring to ``drop out''.
This is a problem even if peers are initially identified and
authenticated prior to joining because peers may be compromised and
an initially nonmalicious peer may become malicious.

In fact, the only assumption that we can reasonably make, assuming
that all peers are identified and authenticated before joining, is
that only a small fraction of peers are malicious.  The challenge
then, is to limit the ability of the malicious peers to collaborate
and disrupt the overlay, to detect malicious peers, and evict them 
from the overlay.

\section{Design and Implementation}\label{sec:imp}
Our implementation is an extension of the I3~\cite{StAdZhShSu04}
code-base.  Our implementation comprises five parts: (i) a key
assignment scheme that links each peer's key with its network
address\footnote{Both IP address and port number.} while at the
same time uniformly distributing the peers' keys in the ring; (ii)
a distributed identification, authentication, and (revocable)
authorization mechanism that allows the overlay to control what
peers are admitted into the ring; (iii) a message authentication
mechanism that links each message to its sender; (iv) a self-policing
mechanism based on overlapping groups composed of proximate peers;
and (v) a simple form of random routing that avoids the possibility
of any peer becoming a choke point between two other peers.

\subsection{Key Assignment}
As was observed in~\cite{SiMo02} and~\cite{CaDrGaRoWa02}, it is
harder for malicious peers to collaborate when they are uniformly
distributed in the ring than when they are clustered.  Consequently,
peers should be assigned keys from a uniform distribution.  Thus,
prior to joining, each peer is expected to choose a key from the
uniform distribution on the key space.  However, there is nothing
that prevents malicious peers from choosing keys that facilitate
collaboration.  Furthermore, a randomly selected key, only encodes
the peers position within the ring, not the network, which another
peer would need to contact it directly.  Lastly, the choice of the
peer's network address is typically limited and in most cases beyond
the control of the peer, malicious or otherwise.

We leverage this restriction to assign keys to peers so that the
peers have no choice in their key, the key is unique, the key encodes
a peer's network address, and the key appears to be chosen from the
uniform distribution on the key space.  To determine its key a peer
concatenates its IP address and port number, both in network byte
order, to create a 6 byte string.  This string is passed through
the SHA-1 function, generating a 20 byte hash.  The hash is the
same length as a key, 20 bytes, and appears as if it was chosen
from the uniform distribution on the key space.\footnote{In reality
the hash is uniformly chosen from key subspace of cardinality
$2^{48}$, the size of the input string.}  Lastly, the IP address
and the port number replace the 6 least significant bytes of the
hash, as suggested in~\cite{CaDrGaRoWa02}.

The resulting 20 byte key, can easily be validated by extracting
the 6 least significant bytes, passing them from the SHA-1 function,
and comparing the 14 most significant bytes of the resulting hash
and the key---they should match.  The 14 most significant bytes of
the key look as if they were drawn from a uniform distribution,
ensuring that the peers are uniformly distributed throughout the
ring.  Lastly, the key uniquely identifies each peer because the
IP address of each peer is necessarily unique.  Thus, each peer 
can be uniquely identified.

\subsection{Distributed Identification, Authentication and Authorization}
A peer must be identified, authenticated, and authorized before it can
join the overlay.  The peer's key uniquely identifies the peer, but it
does not authenticate the peer, which is a prerequisite for authorization.  
Since the maliciousness of a peer may be discovered only after it joins 
the ring, authorization must be revocable, in order to facilitate the 
excommunication of such peers.

Authentication is accomplished by using a public key signature
system---each new peer generates a public-private key-pair.  A peer
authenticates a message by first embedding its 20-byte key into the
message and then signing it.  However, two problems remain:
distribution of the public key, and the authorization of the peer.
Both of these problems are solved simultaneously by leveraging the
Domain Name System (DNS)~\cite{rfc1034,rfc1035}.

Each ring is identified by a domain name in the DNS database and
each authorized peer in the ring has corresponding a TXT entry
within the domain, identified by the peer's key and storing a
certificate that contains the peer's public key.  The authority
responsible for authorizing peers is also responsible for signing
the certificates and for adding or removing the TXT entries.

When a peer receives a message from another peer, it checks its
cache for the sender's public key, if present then the sender is
authorized to participate in the ring.  Otherwise, the receiver
performs a DNS lookup for the sender's key in the ring's domain.
If found, the sender's public key is added to the cache and the
sender is deemed to be authorized.  If not, a negative entry is
added to the cache, causing the peer to ignore all future messages
from the sender until the negative entry expires.  Authorizations
are revoked by removing the corresponding TXT entry from the DNS
database and informing all peers via a broadcast.

We leverage the DNS system because it has proven to be relatively
robust and fault tolerant.  In fact, robustness can be increased
by simply adding more name servers.  Furthermore, a DNS query is
only needed when a new peer joins.  In theory, peers could broadcast
the certificates they receive from their DNS queries, informing the
ring of the joining peer.  Thus, an attack on the DNS system would
only prevent new peers from joining the ring.  One problem with our
approach is that authenticating each message using a public key
signature is prohibitively expensive.

\subsection{Message Authentication}
A message is linked to its sender because it contains the sender's
key and then signed by the sender.  Since the keys are unique and 
contain the sender's network address, each message can be traced 
to its origin.  Thus, if fraudulent messages are detected, the 
sender can be identified with certainty and excommunicated.

Unfortunately, signing and verifying all messages using a public
key signature system is expensive.  For example, to determine the
overhead of using a public key signature system, we ran a two peer
ring on a single 1.60GHz Intel Xeon E5310 (4-core) server with
2 gigabytes of RAM, and had one peer ping the other.  This nullified
the any potential network related slowdown, and allocated one CPU
to each peer, thus avoiding any issues associated with sharing a
CPU.  Without message authentication, the system performed about
4000 pings per second---approximately 8000 messages per second.
With message authentication, using public key signatures, the number
of pings per second dropped to 15---a slowdown by a factor of 300!

We solve this problem by using message authentication codes (MAC)
as the default authentication mechanism.  The Chord overlay structure
exhibits good temporal locality with respect to communication,
meaning that if a peer communicates directly with another peer, it
will do so repeatedly in the future.  The first time two peers
communicate directly, they exchange shared secret keys (using public
key encryption), and use shared keys to authenticate all messages
to each other.  Using HMAC based authentication, the performance of
our system went back up to about 3500 pings per second.

\subsection{Our Brothers' Keepers}
Chord overlay structure relies on peers behaving properly: forwarding
requests that they cannot satisfy and replying truthfully to requests
that they can satisfy.  However, if a malicious peer does not forward
requests, or even worse, misdirects the requests or sends fraudulent
replies, the overlay structure can be subverted.  In particular,
maligning the ``find successor'' requests, which are used by peers
to find their position within the ring and construct finger tables,
can create loops and partitions within the ring, rendering the
overlay dysfunctional.  That is, a few collaborating malicious peers
could cause segments of the ring to ``drop out''.

Realistically, we can neither ensure that no malicious peer will
ever join, nor can we ensure that no peer will ever be compromised.
Malicious peers are distinguished by their behaviour that, when
detected, can be quashed by excommunicating the peer.  Thus, by
increasing the system's ability to detect malicious behaviour, the
amount of damage caused by a malicious peer can be limited.  Since
our key assignment scheme ensures that with high probability two
malicious peers will not be near each other in the ring, we use a
peer group approach to improve detection of malicious behaviour,
i.e., the peer's proximate peers keep it honest.

Each peer in the ring, is associated with a peer group of size~$g$,
where~$g$ is a small odd number, such as $5$, $7$, $9$, $11$, etc.
The group comprises the peer itself---the group leader---and $g -
1$ of its closest peers: $\frac{g - 1}{2}$ closest preceding peers
and $\frac{g - 1}{2}$ closest succeeding peers.  Thus, each peer
belongs to $g$ overlapping groups of size $g$.   Furthermore, given
our assumption about the uniform distribution of malicious peers,
the chance of a group having multiple malicious peers is small.

When a new peer joins the ring, it queries its predecessor and
successor for their group memberships, constructs its own group
membership list from the responses, and then queries the other peers
in its group to confirm their membership.  On an ongoing basis, the
peers in a group query each other's membership lists, updating them
as peers join or leave.  In closed overlays, particularly in the
case of Trinity, we assume that the rate at which peers join and
leave the ring is relatively low.  Hence, a peer's group membership
list will not change often.

In fact, a peer is only added to a group only after it has been 
verified by the group's leader, ensuring that group lists only 
contain valid peers.  These group lists also provide a fast 
mechanism for finding a new successor or predecessors if the 
current one leaves (or fails) the ring.

A peer's group membership list, should be consistent with those of
the group's members, e.g., if the group of peer $p$ is $(n,o,p,q,r)$,
then peer~$q$'s group should be $(o,p,q,r,s)$.  Thus, if a peer
sends a group list that is inconsistent with the lists of other
group members, it is considered malicious, or at least untrustworthy.
Consequently, malicious peers cannot easily send fraudulent ``find
successor'' responses about their group members, because similar
queries to their neighbours would unmask them.  The result is that
peers cannot send out false ``find successor'' replies to any of
its neighbouring peers without being excommunicated.

However, it is also necessary to ensure that remote peers are also
honest, i.e., those peers that are not within a peer's group.  This
is accomplished by leveraging the group structure.  Specifically,
a peer's ``find successor'' response is be verified by querying a
member of its peer group, and is based on the fact that peers in the 
same group will have similar finger tables.  

Recall, that a peer's $i$th finger table entry contains the successor
to key $k + 2^i \bmod 2^{160}$, where $k$ is the peer's key.  Assuming
that peers are uniformly distributed in the ring, if peers with
keys $k$ and $k'$ are adjacent, then the successors to $k + 2^i
\bmod 2^{160}$ and $k' + 2^i \bmod 2^{160}$ will likely be close
to each other in the ring, if not the same peer.  Thus, there will
be considerable overlap between the groups associated with the $i$th
finger table entries of the two peers.  Consequently, a ``find
successor'' response can be verified by resending the query to a
member of the responder's group.

To facilitate this approach, and to verify the consistency of the
groups associated with the finger table entries, our implementation
uses an expanded finger table that stores the keys of the peer's
entire group rather than just the peer's key---the finger table
stores $g$ keys per entry.  Furthermore, a peer's ``find successor''
response includes the keys of the peer's entire group.  Since ``find
successor'' queries are sent on an ongoing basis, the finger table
entries are updated and checked on a regular basis.  Lastly, storing
entire groups in the finger table, instead of single peers, facilitates
the implementation of a simple randomized routing scheme, mitigating
the problem of packet dropping by malicious peers.

\subsection{Randomized Routing}
Even if a malicious peer does not send fraudulent routing responses,
it can still cause problem by simply dropping all messages.  If a
malicious peer is a choke-point between two other peers---all
messages from one peer to the other are routed through it---then
none of the messages may get through.  Detecting this behaviour is
problematic because the I3 Chord implementation and many other
overlay systems use lightweight connectionless unreliable transport
protocols, such as UDP.  Consequently, it is impossible to distinguish
between poor network connectivity and a misbehaving peer.  Fortunately,
our scheme can mitigate both problems.  We note that we cannot
ensure that no messages will be lost; only that with high probability,
not all the messages will pass through the same peer, while in
transit.

We use a variant of randomized routing \cite{LeMaRaRa94}.  Traditional
randomized routing forwards the message to a randomly chosen peer
in the system, and then from that peer to the destination.  This
can dramatically increase the latency, particularly if the destination
peer is close to the sender but the randomly chosen peer is far
away.  Instead, in our scheme, multiple messages between two peers
take different but comparable length paths, ensuring that a
choke-point can not form.

When a message arrives at a peer, the peer classifies the message's
destination as either local, near, or far.  If the destination is
local, then the message has arrived at its destination.  If the
destination is near, then the message is destined to a neighbour
of the peer and is forwarded directly to its destination.  Otherwise,
a peer is selected and the message is forwarded to it.

According to the traditional deterministic forwarding protocol, the
peer whose key most closely precedes the message destination is
chosen from the finger table, and the message is forwarded to this
peer.  In our implementation, a group is chosen from the finger table
such that the group leader's most closely precedes the message
destination.  Then, a peer is randomly chosen from this group and
the message is forwarded to it.  Since the finger tables of the
peers in a group are similar, the route taken between two peers
will differ in the peers that the messages transit.  However, as
discussed in the preceding section, these peers are near each other
within the ring, implying that the total number of hops will not
vary greatly.

The correctness of the protocol does not change as long as the key 
of the peer selected from the finger table precedes the message
destination, and since all peers in a group are, by definition,
near each other, the size of each hop is will differ by an 
additive constant, resulting in a small variance in the number
of hops that a message takes.

\section{Evaluation}\label{sec:eval}
To evaluate the performance of our implementation we used a 255
peer ring running on a 26 machine cluster running OpenBSD 4.3 and
4.2.  One of these machines was an Intel Xeon X3210 2.13GHz Quad-core
based server with 4GB of RAM, which ran 5 peers on it and served
as the name server for the cluster.  Each of the remaining 25
machines was an Intel Pentium 4 2.80 GHz based desktop with 1 GB
of RAM.  Each of these desktops ran 10 peers each and all the
machines were interconnected via a Cisco WS-C2924--XL-EN and a Cisco
WS-C3548-XL-EN managed switches that were locked at 10 Mb/s
half-duplex---the mean latency between any two machines in the
cluster was 0.5 milliseconds, with a negligible variance.
We performed several different tests to measure the latency,
throughput, and capacity of our implementation in both secure and
insecure modes, in order to compare the overhead associated with 
secure mode.

\subsection{Latency and Throughput}
We first compared the latency and throughput overhead of secure
versus insecure operation.  Since peers regenerate and exchange
their shared keys at regular intervals, different parts of ring had
different loads at different times.  To compensate for this, a
series of test runs were performed, spanning a sufficiently large
time interval, and the minimums over these test runs were used.

Each test comprises two communicating peers: the initiator, which
conducts and times the test, and the responder, which serves as the
other end-point of the communication.  The latency test measures
the round trip time of a ping and its echo.  The initiator pings
the responder, which echos the ping---both the ping and the echo
are routed through the overlay.   The test is repeated sequentially
a set number of times and the count is divided by the total time,
yielding the round trip time per ping.  The throughput test measures
how fast packets (or messages) can be sent through the overlay.
The initiator sends a throughput request to the responder, indicating
the number of packets the responder should send back.  The responder
sends the requested number of packets (through the overlay) as
quickly as possible, and the initiator measures the time difference
between the arrival of the first and last packets---the number of
packets divided by the difference is the throughput.

Our evaluation fixed one of the five peers on the 4-core server to
be the initiator, and used the 250 peers running on the 25 desktops
as responders.  For both latency and throughput measurements, the
initiator performed 12 test series consisting of 10 test runs that
consisted of 250 tests, once for each peer.  Each latency test
performed 10 pings at a time and each throughput test had the
responder send back 1000 packets.  Each series takes the minimum
measurement for each peer over the 10 runs.  The minimums for each
peer from the 12 series are averaged to yield the latency or
throughput measurement.

Table~\ref{tab:summary} displays the mean, median, maximum, minimum,
and standard deviation round trip times and throughput measured
for all 250 peers.  The table shows the measurements for both
insecure mode operation and secure mode operation, and the overhead
of the secure mode.

\begin{table}[htb]
\begin{center}
\begin{tabular}{|l|l|l|l||l|l|l|}
\cline{2-7} 
\multicolumn{1}{c}{\ }
& \multicolumn{3}{|c||}{Latency} & \multicolumn{3}{|c|}{Throughput} \\
\cline{2-7} 
\multicolumn{1}{c|}{\ }
& Insecure Op. & Secure Op. & Relative & Insecure Op. & Secure Op. & Relative \\
\multicolumn{1}{c|}{\ }
& RTT (sec) & RTT (sec) & Difference & Pkts / sec & Pkts / sec & Difference \\
\hline
Mean    & 0.002874 & 0.003457 & 20.2\% & 6148 & 4946 & 19.4\% \\
Median  & 0.002897 & 0.003483 & 20.2\% & 6389 & 5087 & 20.4\% \\
Maximum & 0.003542 & 0.004282 & 20.9\% & 7794 & 6566 & 15.8\% \\
Minimum & 0.000759 & 0.000880 & 15.9\% & 3107 & 2643 & 14.9\% \\
Std. Dev.   & 0.000335 & 0.000411 & N/A    & 1164 & 930 & N/A \\
\hline
\end{tabular}
  \caption{Summary statistics of round trip times to peers and packets per 
           second from peers.\label{tab:summary}}
\end{center}
\end{table}

The measured latency in secure mode is 20\% greater than the latency
in insecure mode.  Although, this seems high, it is important to
remember that there were 10 peers running on each host, making the
system CPU bound and that the time difference, 0.6 milliseconds, is negligible
compared to the typical latency between two hosts in the Internet.

The throughput in secure mode is also on average 20\% lower.  This
is due to the cost of authenticating messages: the sender has to
sign each message and the receiver has to verify the message.  Since
message authentication is a CPU bounded task, its effect will be
less when only one peer is running on each server.

\begin{figure*}[htb]
  \begin{center}
  \ \mbox{
  \includegraphics[width=0.48\linewidth]{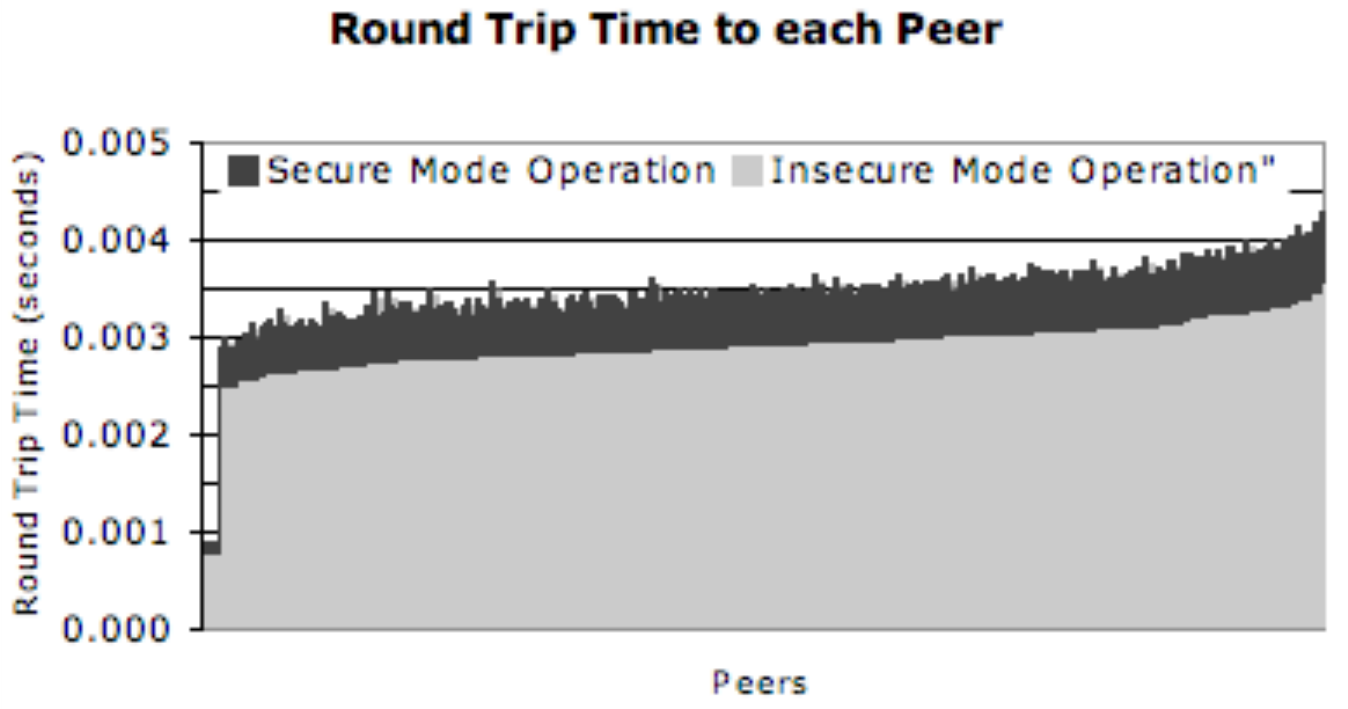}\ 
  \includegraphics[width=0.5\linewidth]{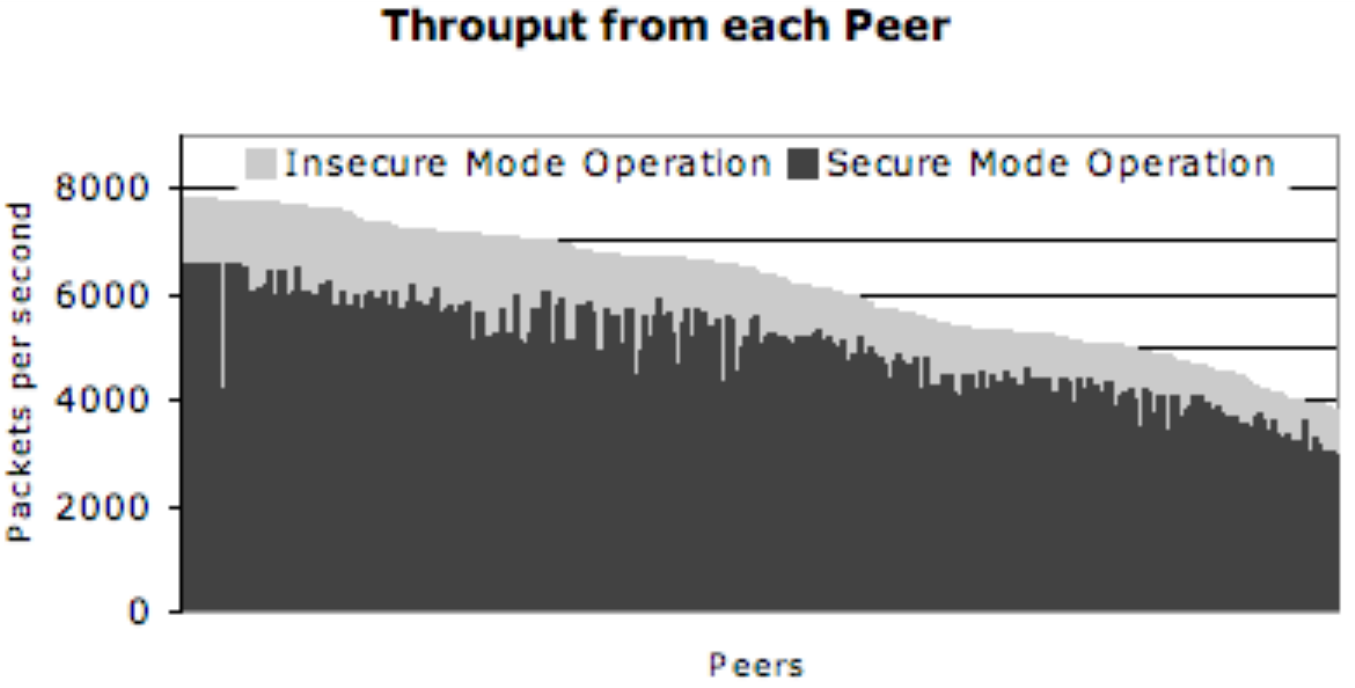}}\ 
  \caption{Round trip times to peers.\label{fig:pings}}
  \end{center}
\end{figure*}

It is more instructive to view the round trip times for each peer
and throughput from each peer in a sorted order.  The first graph
in Figure~\ref{fig:pings} shows the round trip times to all the
peers for both insecure and secure operation modes, in ascending
order of times measured in insecure mode.  The second graph in
Figure~\ref{fig:pings} shows the throughput from all the peers for
both insecure and secure operation modes, in descending order of
times measured in insecure mode.

Several artifacts are immediately visible in the first graph: First,
four peers have much lower round trip times.  These peers are the
successors and predecessors of the peer performing the ping, and
hence both the ping and the response only take one hop.  Second,
there is large jump in round trip times for both insecure and secures
modes; approximately, 0.0025 and 0.003 seconds respectively.  Since
the minimum latency between two peers in the cluster is 0.0005
seconds, this means that pings to and from all the other peers take
between 6 and 9 hops, which makes sense for a ring of 255 peers.
Lastly, and most importantly, the relative difference in latency
between insecure and secure operation remains fixed, at 0.06
milliseconds per hop.

The second graph also exhibits a couple important features.  First,
the graph has a step feature, corresponding to the distances between
the initiator and the responders.  The closer a responder is to the
initiator, the higher the measured throughput.  Second, the relative
decrease in throughput between insecure and secure operation remains
relatively constant.  As before the primary reason for the reduction
is the cost of message authentication and is noticeable because 
10 peers were running on each singe-core machine.

\subsection{Capacity}
The capacity of an overlay is the measure of the number of messages
that the system can deliver per unit time.  To measure the system's
capacity we implemented a game of hot-potato over the overlay: A
set number of messages (potatoes) are injected into the system.
The potatoes are randomly passed from peer to peer, and counter in
each potato tracks the number of times the potato is passed.  By
varying the number of concurrent potatoes in the system, we control
the system's load.

When a peer receives a potato, it increments the potato's counter,
generates a random key, and sends the potato to the peer responsible
for the random key.  To ensure that no potato is dropped, the
receiving peer acknowledges the potato, and the sender acknowledges
the acknowledgment.  Only after receiving the second acknowledgment
does the receiver commence the next potato pass.  If potato's
originator receives it, and the potato has been in the system for
a minimum amount of time, e.g., 60 seconds, the number of passes
per second for the potato is computed, by dividing the value of the
potato's pass counter by the number of seconds that the potato was
in the system.  The potato's time to live counter is then decremented,
and if nonzero, the potato's pass counter is reset and the potato
is injected into the system again.  This ensures a period of
consistent load.

In each of the runs, the first measurement from the first 75 ejected
potatoes was used.  Table~\ref{tab:passes} exhibits the mean, median,
standard deviation, maximum, and minimum number of passes per second
that a potato achieved under different system loads:  10, 20, 30,
40, 50, 60, 70, 80, and 160 potatoes in the system.  Note: a pass
consists of a 3-message exchange between two peers in the system
and message delivery may take multiple hops within the overlay.

\begin{table}[htb]
\begin{center}
\begin{tabular}{|l|l|l|l|l|l|l|l|l|l|}
\hline
\# of msgs & 10   & 20   & 30   & 40   & 50   & 60   & 70   & 80   & 160 \\
\hline
\hline
\multicolumn{10}{|c|}{Insecure Mode Operation} \\
\hline
Mean       & 163  & 134  & 107  & 86   & 72   & 60   & 51   & 45   & 23 \\
Median     & 163  & 134  & 106  & 86   & 72   & 60   & 51   & 45   & 22 \\
Std. Dev.  & 3.3  & 2.8  & 2.1  & 2.3  & 2.0  & 1.6  & 1.4  & 1.8  & 2.4 \\
Maximum    & 172  & 141  & 113  & 92   & 77   & 65   & 54   & 50   & 33 \\
Minimum    & 156  & 127  & 103  & 81   & 67   & 56   & 48   & 42   & 19 \\
\hline
\hline
\multicolumn{10}{|c|}{Secure Mode Operation} \\
\hline
Mean       & 138  & 115  & 93   & 76   & 62   & 53   & 45   & 40   & 20 \\
Median     & 137  & 115  & 94   & 76   & 62   & 53   & 45   & 39   & 19 \\
Std. Dev.  & 3.4  & 2.0  & 2.1  & 1.6  & 1.4  & 1.3  & 1.6  & 1.8  & 2.6 \\
Maximum    & 147  & 120  & 98   & 79   & 68   & 55   & 49   & 46   & 29 \\
Minimum    & 131  & 109  & 89   & 71   & 58   & 47   & 42   & 36   & 15 \\
\hline
\end{tabular}
  \caption{Number of passes per second that a message takes. \label{tab:passes}}
\end{center}
\end{table}

\begin{figure*}[htb]
  \begin{center}
  \ \includegraphics[width=\linewidth]{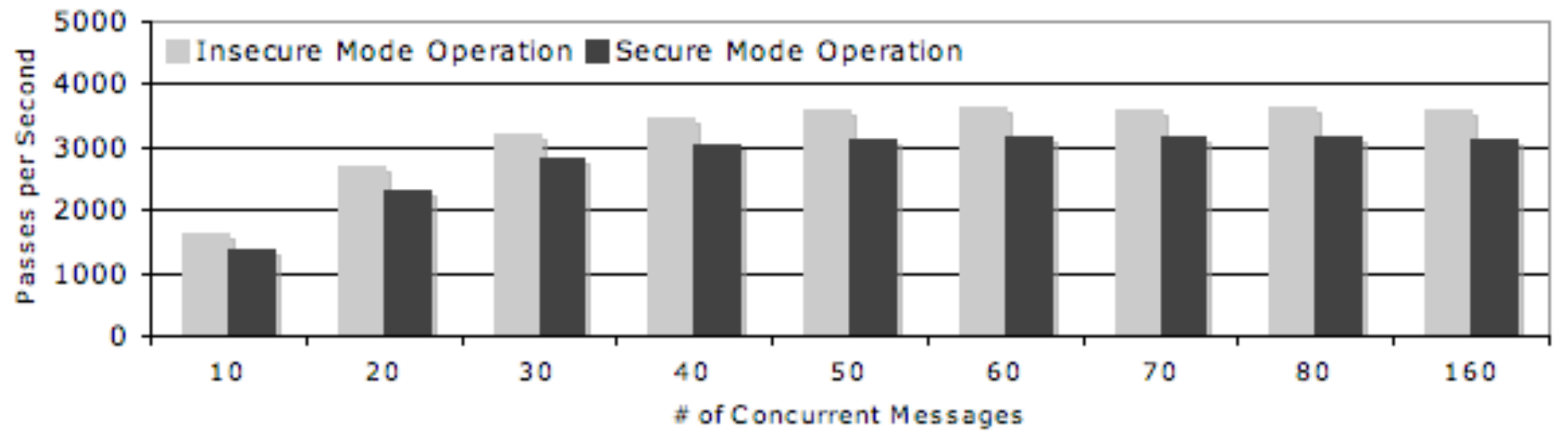}\ 
  \caption{Capacity of overlay.\label{fig:cap}}
  \end{center}
\end{figure*}

As the load increases, the number of passes per second of a potato
decreases because the likelihood that a peer may need to process
multiple potatoes at once increases.   However, passes per second
of a potato does not yield a measure of the capacity of the system
as a whole.  The capacity of the system is the number of passes per
second that the system performs over all.  This is equal to the
average number of passes per second multiplied by the number potatoes
in the system.

Figure~\ref{fig:cap} exhibits the capacity of the system for both
insecure and secure operation modes.  The capacity of the system
is 3600 and 3150 passes per second in insecure and secure operation
modes, respectively.  In both cases the system becomes saturated
at 50 potatoes, but capacity does not degrade as the number of
potatoes increases.  The relative difference in capacity is 12.5\%, 
and is predominately affected by the CPU bounded task of message 
authentication.

\section{Related Work}\label{sec:related}
The challenge of securing peer-to-peer systems has been around since
their advent.  Sit and Morris~\cite{SiMo02} first identified a set
of design principles for securing peer-to-peer systems and described
a taxonomy of various attacks against them.  This work was extended
by Wallach~\cite{Wa02} who investigated the security aspects of
systems such as CAN~\cite{RaFrHaKaSh01}, Chord~\cite{StMoKaKaBa01},
Pastry~\cite{RoDr01}, and Tapestry~\cite{ZhKuJo01}, and discussed 
issues such as key assignment, routing, and excommunication of 
malicious peers.

Castro et al.~\cite{CaDrGaRoWa02} proposed several approaches to
securing peer-to-peer overlays.  They proposed to delegate assignment
of keys to trusted certification authorities, that would ensure
that the keys are chosen at random, and that each peer is bound to
a unique key, with the peer's IP embedded in the key.  To securely
route messages, they proposed to use constrained routing tables,
which contain keys from specific locations in the overlay.  In our
case Chord already constrains a key's location within the overlay, 
obviating the need for constrained routing tables.  In fact, our 
self-policing and random routing mechanisms leverage this constraint.

Castro et al.~\cite{CaDrGaRoWa02} also proposed a routing failure
test that tries to determine what nodes are malicious.  Their
approach also sends multiple copies of the message through diverse
routes to ensure message delivery.  Our approach is similar but
less resource intensive.  Our system uses the peer groups to detect
faulty routing information, and to ensure that no peer is a choke-point
between two other peers.  Our system does not attempt to ensure the
delivery of all messages, but instead attempts to ensure that some
messages will be delivered.

Lastly, there are many ways to secure a peer-to-peer system, for
example LOCKSS~\cite{MaRoRoBaGiMu03} uses majority voting replicas
and computationally rate-limiting cryptographic puzzles~\cite{DwNa92}.
Unfortunately, these approaches severely impact system performance
and are not practical in the context where good performance is a
necessity.

\section{Conclusion and Future Work}\label{sec:conc}
We have designed and implemented a secure and efficient extension
to the I3~\cite{StAdZhShSu04} implementation of the Chord structured
overlay~\cite{StMoKaKaBa01}.  Our extension is aimed at closed
overlays in which membership is tightly controlled.  This context
requires mechanisms for peer identification, authentication, and
authorization, mechanisms for message authentication, and mechanisms
to mitigate the behaviour of malicious peers in the overlay, which
are unavoidable.

Our implementation uses a simple hashing scheme to generate keys
that are linked to peer's network address, and are uniformly
distributed in the key space.  The keys are embedded into messages,
linking each message to its sender via an efficient two-part
authentication mechanism, combining public key and HMAC message
authentication.  Secure routing is implemented via self-policing
peer groups that force malicious peers to either behave properly
or face detection.  Lastly, these groups are leveraged for a simple
random routing scheme that prevents choke-points within the overlay.

Our evaluation, which was performed on a local cluster, has
demonstrated that our implementation's overhead, of about 20\%, is
primarily due to CPU bounded operations.  We believe that this
effect will significantly decrease under normal conditions in the
larger Internet context where latency will dominate, and where
multiple peers are not running on the same host.

To validate this hypothesis, we intend to perform a more realistic
evaluation using the Planet-Lab platform, which spans the world and
will allow us to test much larger overlays.  We are in the process
of implementing the Trinity~\cite{BrBr07} e-mail classification
system on top of our secure overlay.  This will provide additional
opportunities to identify and solve performance bottlenecks in our
implementation.

\bibliographystyle{alpha} 
\bibliography{biblio}

\end{document}